\documentclass[sn-mathphys,iicol]{sn-jnl}
\usepackage{bm}
\usepackage{graphicx}

\jyear{2023}

\begin{document}

\title[Efficiency and thermodynamic\ldots]{Efficiency and thermodynamic uncertainty relations of a dynamical quantum heat engine}

\author*[1,2]{\fnm{Luca} \sur{Razzoli}}\email{luca.razzoli@uninsubria.it}
\author[3,4]{\fnm{Fabio} \sur{Cavaliere}}\email{fabio.cavaliere@unige.it}
\equalcont{These authors contributed equally to this work.}
\author[4]{\fnm{Matteo} \sur{Carrega}}\email{matteo.carrega@spin.cnr.it}
\equalcont{These authors contributed equally to this work.}
\author[3,4]{\fnm{Maura} \sur{Sassetti}}\email{maura.sassetti@unige.it}
\equalcont{These authors contributed equally to this work.}
\author[1,2,5]{\fnm{Giuliano} \sur{Benenti}}\email{giuliano.benenti@uninsubria.it}
\equalcont{These authors contributed equally to this work.}
\affil*[1]{\orgname{Center for Nonlinear and Complex Systems, Dipartimento di Scienza e Alta Tecnologia}, \orgdiv{ Universit\`a degli Studi dell'Insubria}, \orgaddress{\street{Via Valleggio 11}, \postcode{22100} \city{Como}, \country{Italy}}}
\affil[2]{\orgdiv{Istituto Nazionale di Fisica Nucleare, Sezione di Milano}, \orgaddress{\street{Via Celoria 16}, \postcode{20133} \city{Milano}, \country{Italy}}}
\affil[3]{\orgdiv{Dipartimento di Fisica}, \orgname{Universit\`a di Genova}, \orgaddress{\street{Via Dodecaneso 33}, \postcode{16146} \city{Genova}, \country{Italy}}}
\affil[4]{\orgdiv{CNR-SPIN}, \orgaddress{\street{Via Dodecaneso 33}, \postcode{16146} \city{Genova}, \country{Italy}}}
\affil[5]{\orgdiv{NEST, Istituto Nanoscienze-CNR}, \orgaddress{\street{Piazza S. Silvestro 12}, \postcode{56127} \city{Pisa}, \country{Italy}}}

\abstract{In the quest for high-performance quantum thermal machines, looking for an optimal thermodynamic efficiency is only part of the issue. Indeed, at the level of quantum devices, fluctuations become extremely relevant and need to be taken into account. In this paper we study the thermodynamic uncertainty relations for a quantum thermal machine with a quantum harmonic oscillator as a working medium, connected to two thermal baths, one of which is dynamically coupled. We show that parameters can be found such that the machine operates both as a quantum engine or refrigerator, with both sizeable efficiency and small fluctuations.}


\maketitle
\section{Introduction}\label{sec:intro}
The rapid development of quantum technologies calls for a deeper understanding of thermodynamics and energetics at a microscopic level, where unavoidable quantum effects have to be taken into account. The extension of classical concepts of thermodynamics to the quantum realm is not only at the forefront of fundamental theoretical research~\cite{campisi11,kosloff12, niedenzu, anders, talkner20, benenti17}, but also relevant for new nanodevice applications~\cite{myers22,bjorn, giazotto,martinez, pekola, arrachea}. Mastering the thermodynamics of quantum systems far from equilibrium, from energy storage to energy transfer and transduction and heat--to--work conversion, is of great importance for emerging technologies with applications in quantum computing~\cite{arrachea,krantz19, calzona}, quantum communication~\cite{aufeve, qcbook}, and quantum sensing~\cite{golokolenov}.

Spurred by the rapid emergence of new quantum technology platforms, in the last few years the first prototype realizations of quantum thermal machines~\cite{pekola, spin, cangemireview, sheng,vischi}  and quantum batteries~\cite{quach, maillette} have been reported, calling for new theoretical and experimental investigations, even including extensions to non-Hermitian quantum thermodynamics~\cite{Gardas_2016,Santos_2021}. For instance, a great interest revolves around heat nano-engines and refrigerators where the working medium (WM) is a quantum system coupled to several thermal reservoirs. Indeed, it has been shown that a multi-terminal configuration can improve both the output power and the efficiency at maximum power~\cite{levy12, huber19, imry12, mazza04,mazza15,erdman17,sanchez2,cavaliere_iscience,lopez23,lu23}. Also, the impact of quantum coherence on the performance of thermal machines has been inspected~\cite{karimi16,pekola19,brandner17,kosloff02,rezek06,fried17,korze16,hammam22,camati_pra_19,liu_pre_21}, as well as the role of the uncertainty principle~\cite{chattopadhyay}.

Quantum devices are highly sensitive to fluctuations that may limit device performances~\cite{anders, cangemireview, paternostro, gomez0, gomez}, in contrast to macroscopic devices, where usually fluctuations can be safely neglected. To assess fluctuations, the thermodynamic uncertainty relations (TURs)
have been recently introduced~\cite{barato, shiraishi16, timpanaro,
guarnieri, asegava, pal, shiraishi18, cangemi,menczel, koyuk, potanina} and a unified toolbox for describing current fluctuations in the context of open quantum systems has been proposed~\cite{landi_arxiv_23}. These TURs combine steady-state currents $J$, their fluctuations $D_{J}$ and dissipation (measured by the entropy production rate $\dot{S}$), 
giving limits on the precision of currents for 
a given dissipation. 
Indeed, in Ref.~\cite{barato} it was reported that for certain classical
Markovian systems, the signal-to-noise ratio satisfies a TUR bound, showing that relative current fluctuations are 
lower bounded in terms of the inverse entropy
production rate. By introducing a dimensionless trade-off parameter $Q_J$, 
TURs can be expressed as 
\begin{equation}
Q_J\equiv \dot{S}\frac{D_{J}}{J^2}\ge 2 k_B,
\end{equation}
where $k_B$ is the Boltzmann constant (from now on we will set $k_B=1$).
After that, several bounds for TURs~\cite{horowitz17, pietzonka18,
koyuk18, brandner18, coherence18,cangemi20,jiao20,singh,miller_prl_21}, and their possible violations, have
been put forward under different assumptions, starting from classical
Markovian dynamics to quantum systems, including time-dependent forces
and the role of time-reversal symmetry.
For instance, when the model preserves time--reversal--symmetry it
has been shown that, at least in the linear regime~\cite{cangemi,
brandner18}, the trade-off parameter $Q$ 
never achieves values lower than 2. Recently, experimental results on TURs have appeared~\cite{pal}. They thus
represent new tools to inspect the performances of so-far unexplored
resources based on non-thermal states, correlations and quantum
coherences in non-equilibrium quantum systems~\cite{paternostro}.

In this work, we propose a physically implementable quantum thermal machine and investigate its operation. We consider a single quantum harmonic oscillator, a paradigmatic model and a building block of many quantum technologies~\cite{krantz19, arrachea,cavop1,cavop2,cqed1,zherbe95,blais_pra04,barzanjeh2,pontin}, as the WM coupled to two thermal reservoirs. One of them is statically linked, while the other one is modulated at a frequency $\Omega$ (monochromatic driving). Driven system-bath couplings offer enhanced design flexibility, such as the possibility to switch between regimes of quantum heat engine or refrigerator~\cite{cavaliere_prr,cavaliere_iscience}. Here we are interested in assessing the precision of the currents in such regimes, by studying the TURs related to the heat current and the 
work current (i.e., the total exchanged power). Via a systematic perturbative approach, we obtain expressions for the thermodynamic quantities and the TURs, that can be numerically integrated with standard techniques. Even though we report no violation of TURs, i.e., we do not observe values of $Q < 2$, we show that $Q$ can attain minimal values (close to 2) both in the engine and refrigerator regimes, accompanied by sizeable efficiency.
Values $Q \approx 2$ are related to a nearly optimal trade-off of the three desiderata we commonly want for a heat engine (refrigerator)---finite output power (cooling power), efficiency (coefficient of performance) close to the Carnot value, and small fluctuations---as $Q$ can be explicitly written in terms of them~\cite{pietzonka18}. In the limit of very small damping, simple analytical expressions for the TURs are obtained, which support our findings.

In Sec.~\ref{sec:modther} the model is illustrated, along with the definition of all thermodynamic quantities and corresponding correlators describing their fluctuations. We then evaluate their quantum and time averaged values in the dynamical steady state, to lowest order in the strength of the dynamical coupling. Results are reported and discussed in Sec.~\ref{sec:sec4}, and conclusions are drawn in Sec.~\ref{sec:conc}. Appendix~\ref{app:pert} provides details on the adopted perturbative approach, while Appendix~\ref{app:opmodes} illustrates all the possible operating modes of the proposed quantum thermal machine.

\section{Model and general setting}\label{sec:modther}
\subsection{Model}
\begin{figure}
    \centering
    \includegraphics[width=7.5cm,keepaspectratio]{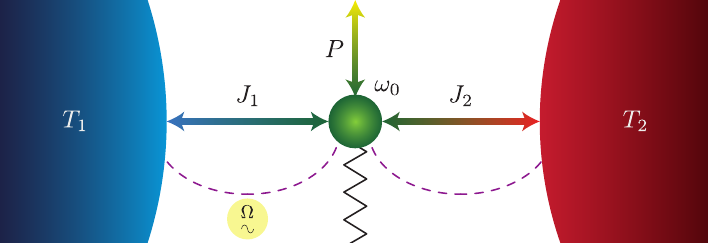}
    \caption{{\bf Setup}. Schematic depiction of the setup under investigation. The WM -- a quantum harmonic oscillator with characteristic frequency $\omega_0$ -- is in contact with two thermal reservoirs at temperatures $T_\nu$, with $\nu =1,2$. The WM can exchange heat currents $J_\nu$ with the reservoirs and total power $P$ with an external source. The purple dashed lines represent the coupling between the WM and the thermal reservoirs. The coupling to the $\nu = 1$ reservoir is modulated in time with frequency $\Omega$, while that to the $\nu = 2$ reservoir is static.}
    \label{fig:setup}
\end{figure}
We consider a minimal model for an operating heat engine in the quantum regime. A single quantum harmonic oscillator acts as the WM and is coupled to two thermal reservoirs, respectively kept at temperatures $T_1$ and $T_2$ --see Fig.~\ref{fig:setup}. The total Hamiltonian is ($\hbar = k_B = 1$)
\begin{equation}
H^{(t)} =  H_{{\rm WM}} + \sum_{\nu=1}^{2} \left[H_\nu + H_{{\rm int}, \nu}^{(t)}\right]\,,
\label{eq:totH}
\end{equation}
where the Hamiltonian of the WM is
\begin{equation}
H_{{\rm WM}}= \frac{p^2}{2 m} + \frac{1}{2}m \omega_0^2 x^2\,,
\label{eq:Ham_qho}
\end{equation}
with $m$ and $\omega_0$ the mass and the characteristic frequency. The Hamiltonian of the $\nu=1,2$ reservoir is described in the usual Caldeira-Legget framework~\cite{weiss, CL83} in terms of an infinite set of independent harmonic oscillators as
\begin{equation}
H_{{\nu}}=  \sum_{k=1}^{+\infty} \left[\frac{P^2_{k,\nu}}{2 m_{k,\nu}} + \frac{1}{2}m_{k,\nu} \omega^2_{k,\nu }X^2_{k,\nu}\right]\,,
\label{eq:Ham_res}
\end{equation}
and the interaction between the WM and the $\nu$--th reservoir is
\begin{equation}
\!\!\!\!\!\!\!\!H_{{\rm int},\nu}^{(t)}\!\!=\!\!\sum_{k=1}^{+\infty} \left[
- g_\nu(t) c_{k,\nu} x X_{k,\nu}
+\frac{g^2_\nu(t) c^2_{k,\nu}}{2 m_{k,\nu}\omega_{k,\nu}^2} x^2\right]\!, 
\label{eq:H_int_nu}
\end{equation}
where $c_{k,\nu}$ describes the interaction strength and the counter--term $\propto x^2$ prevents the renormalization of the WM potential. Following Refs.~\cite{carrega_prxquantum, cavaliere_prr, cavaliere_iscience}, we focus on a dynamical coupling situation to establish a working thermal machine in a periodic steady state regime. In particular, we assume here that the dimensionless coupling to the $\nu=1$ reservoir is modulated in time as $g_1(t)=\cos(\Omega t)$, with $\Omega$ the driving frequency, while the $\nu=2$ coupling is kept constant $g_2(t)=1$ -- see the sketch in Fig.~\ref{fig:setup}. At the initial time $t_0{\to -\infty}$, the reservoirs are assumed in their thermal equilibrium at temperatures $T_\nu$, with the total density matrix written in a factorized form as 
\begin{equation}
\rho(t_0) = \rho_{{\rm WM}}(t_0)\otimes \rho_{1}(t_0)\otimes \rho_{2}(t_0)\,,
\label{eq:initial_rho}
\end{equation}
with
\begin{equation}
    \rho_\nu (t_0) = \frac{e^{-H_\nu /T_\nu}}{{\rm Tr} \left\{ e^{-H_\nu/T_\nu}\right\} }
\end{equation}
and with $\rho_{{\rm WM}}(t_0)$ the initial density matrix of the WM.

In the Heisenberg picture the equations of motion for the WM and reservoir degrees of freedom~\cite{carrega_prxquantum} are, respectively,
\begin{eqnarray}
\dot{x}(t)&=&\dfrac{p(t)}{m},\nonumber\\
\dot{p}(t)&=&- m \omega_0^2 x(t) +{\displaystyle \sum_{\nu=1}^{2}} {\displaystyle \sum_{k=1}^{+\infty}} \bigg[
 g_\nu(t) c_{k,\nu} X_{k,\nu}(t)\nonumber\\
&-&\dfrac{g^2_\nu(t) c^2_{k,\nu}}{m_{k,\nu}\omega_{k,\nu}^2}x(t) \bigg],
\label{eq:EoM_DoF_sys}
\end{eqnarray}
and
\begin{eqnarray}
\!\!\!\!\dot{X}_{k,\nu}(t) &=&\frac{P_{k,\nu}(t)}{m_{k,\nu}},\nonumber\\
\!\!\!\!\dot{P}_{k,\nu}(t) &=&\!\!- m_{k,\nu} \omega_{k,\nu}^2 X_{k,\nu}(t) +  
g_\nu(t) c_{k,\nu} x(t),
\label{eq:EoM_DoF_res}
\end{eqnarray}\\
where overdots denote time derivatives.

The solution for the position operator of the $k$--th oscillator of the $\nu$--th reservoir is
\begin{eqnarray}
\!\!\!\!\!\!X_{k,\nu}(t)&=&\xi_{k,\nu}(t) +\frac{c_{k,\nu}}{m_{k,\nu}\omega_{k,\nu}}\cdot\nonumber\\
&\cdot&\int_{t_0}^t \mathrm{d}s\ g_\nu(s) x(s) \sin[\omega_{k,\nu}(t-s)]\, ,
\end{eqnarray}
\begin{eqnarray}
\!\!\!\!\!\!\xi_{k,\nu}(t)&=&X_{k,\nu} (t_0) \cos [\omega_{k,\nu}(t-t_0)]\nonumber\\
&+&\frac{P_{k,\nu}(t_0)}{m_{k,\nu} \omega_{k,\nu}}\sin[\omega_{k,\nu}(t-t_0)]\,.
\label{eq:xi_knu_def}
\end{eqnarray}
The usual fluctuating force operator, a stochastic force which the reservoir exerts on the WM and that obeys stationary Gaussian statistics, is defined as~\cite{weiss}
\begin{equation}
\xi_\nu(t) \equiv \sum_{k=1}^{+\infty} c_{k,\nu}\xi_{k,\nu}(t),
\label{eq:xi_nu_def}
\end{equation}
with zero quantum average $\langle\xi_\nu(t)\rangle ={\rm Tr}\{\xi_\nu(t)\rho(t_0)\}=0$ and correlation function $\langle\xi_\nu(t)\xi_{\nu'}(t')\rangle \equiv {\cal L}_\nu(t-t')\delta_{\nu,\nu'}$, with
\begin{equation}
\label{corr}
{\cal L}_\nu(t-t')={\cal L}_{\nu,{\rm s}} (t-t') + {\cal L}_{\nu,{\rm a}}(t-t')\,.
\end{equation}
Here, the symmetric (s) and anti-symmetric (a) contributions with respect to the time argument are
\begin{eqnarray}
\!\!\!\!\!\!{\cal L}_{\nu,{\rm s}}(t)&\!\!=&\!\!\!\!\int_0^{+\infty}\!\!\frac{\mathrm{d}\omega}{\pi} 
{\cal J}_\nu(\omega)\coth\left(\frac{\omega}{2T_\nu}\right)\cos(\omega t),\label{eq:ls}\\
\!\!\!\!\!\!{\cal L}_{\nu,{\rm a}}(t)&\!\!=&\!\!-i\int_0^{+\infty}\!\!\frac{\mathrm{d}\omega}{\pi} 
{\cal J}_\nu(\omega)\sin(\omega t).\label{eq:la}
\end{eqnarray}
In the above expressions, the so-called spectral density~\cite{weiss}
\begin{equation}
\label{spectral}
{\cal J}_\nu(\omega)= \frac{\pi}{2}\sum_{k=1}^{+\infty} \frac{c^2_{k,\nu}}{m_{k,\nu}\omega_{k,\nu}}\delta(\omega - \omega_{k,\nu})
\end{equation}
has been introduced. It governs the properties of the $\nu$--th reservoir, e.g. its possible non--Markovian behaviour~\cite{nm1, nm2, nm3, nm4, nm5}. The precise shape of ${\cal J}_{\nu}(\omega)$ will be specified later.
\subsection{Thermodynamic quantities}\label{sec:sec22}
In this work we focus on thermodynamic quantities in the long time limit, when a periodic steady state has been reached. Therefore, all quantities of interest can be averaged over one period of the drive ${\cal T}=2\pi/\Omega$, and are therefore well-defined both for weak and strong coupling~\cite{jurgen,esposito19,liu22, brandner166}. The average heat current $J_{\nu}$ of the $\nu$--th reservoir and the total power $P$ can be written as~\cite{cavaliere_prr}
\begin{eqnarray}
P&=&\frac{1}{\cal T}\int_{\bar{t}}^{\bar{t}+{\cal T}}\!\!\!\!\mathrm{d}t' \sum_{\nu=1}^{2}{\rm Tr}
\left[\frac{\partial H_{{\rm int},\nu}^{(t')}}{\partial t'}\rho(t')\right]\,,\label{eq:P}\\
J_\nu&=&-\frac{1}{\cal T}\int_{\bar{t}}^{\bar{t}+{\cal T}}\!\!\!\!\mathrm{d}t' {\rm Tr}
\left[H_{\nu}\frac{\mathrm{d}}{\mathrm{d}t'}\rho(t')\right]\,,\label{eq:J}
\end{eqnarray}
where $\rho(t)$ is the total density matrix, describing the WM plus reservoirs at time $t$, and, here and in the following, $\bar{t}$ denotes a large positive time. It is worth noting that a positive sign in the above quantities indicates an energy flow towards the WM. Since the power contribution is associated to the temporal variation of the interaction term, it is only due to the dynamical coupling of the WM to the $\nu=1$ bath. An energy balance relation holds true,
\begin{equation}
P +J_1 +J_2 =0,
\label{balance}
\end{equation}
reflecting the first law of thermodynamics~\cite{cavaliere_prr, benenti17, paternostro}. An important quantity to assess the thermodynamic performances is the time--averaged entropy production rate, related to the heat currents by~\cite{benenti17, paternostro, esposito19}
\begin{equation}
 \label{entropy}
\dot S=-\sum_{\nu=1}^{2}\frac{J_\nu}{T_\nu}\,,
\end{equation}
with $\dot{S}\geq 0$ in accordance to the second law of thermodynamics~\cite{paternostro}.

Due to dissipation, all quantities undergo fluctuations that affect the device performances.
During the whole time interval $t-t_0$, fluctuations can be characterized by auto--correlation functions~\cite{weiss, cangemi}: Given an observable $O(t)$ we define 
\begin{equation}\label{eq:meanfluct}
D_{O} (t)=\frac{1}{t-t_0}\int_{t_0}^{t}\mathrm{d}s \int_{t_0}^{t}\mathrm{d}s' \langle{O (s) O (s')}\rangle,
\end{equation}
 where $t-t_0$ is a large time interval, $t-t_0\to+\infty$, and where $\langle A(t)\rangle\equiv\mathrm{Tr}\{A(t)\rho(t_0)\}$ denotes the quantum average. In this limit, the above auto--correlation function reduces to a single integral, that can be written as~\cite{cangemi}
\begin{equation}
\label{doot}
D_{O}(t)=\int_{0}^{+\infty}\mathrm{d}\tau \langle\{O (t), O (t-\tau)\}\rangle\,,
\end{equation}
with $\{A,B\}=AB+BA$ the anti--commutator, leading to the period--averaged fluctuations
\begin{equation}
D_{O}=\frac{1}{{\cal T}}\int_{\bar{t}}^{\bar{t}+{\cal T}}\ \mathrm{d}s\ D_{O}(s).\label{corrav}
\end{equation}

\section{Results}\label{sec:sec4}
The dynamics of the driven dissipative quantum system can be solved by resorting to non-equilibrium Green function formalism~\cite{paz1, paz2, arrachea12}, as described in Ref.~\cite{carrega_prxquantum}. In this work, however, we will only focus on the case in which the modulated coupling ($\nu=1$) is much weaker than the static one ($\nu=2$). Under these circumstances, a perturbative expansion can be used to evaluate all the quantities introduced above.
To do so, it is convenient to rewrite the total Hamiltonian in Eq.~(\ref{eq:totH}) as 
\begin{equation}
H^{(t)} = H^{(0)} + \Delta H^{(t)}\,,\label{eq:totH_pert}
\end{equation}
as the sum of an unperturbed Hamiltonian, $H^{(0)}$, and $\Delta H^{(t)}$, the time--dependent interaction with the reservoir $\nu = 1$:  
\begin{equation}
H^{(0)} =  H_{{\rm WM}}  + \sum_{\nu =1}^{2} H_\nu  + H_{{\rm int},2}\,;\,\Delta H^{(t)} = H_{{\rm int}, 1}^{(t)}.\nonumber
\end{equation}
Deferring all details to Appendix~\ref{app:pert}, here we quote the final results for the quantities of interest defined in Sec.~\ref{sec:sec22} and obtained at the lowest-order perturbative correction.
The total power $P$ and the heat current $J_1$ are
\begin{equation}
P=-\Omega\int_{-\infty}^{+\infty}\frac{\mathrm{d}\omega}{4\pi m} {\cal J}_1(\omega+\Omega)N(\omega)\chi''_0(\omega),\label{eq:pint}
\end{equation}
\begin{equation}
\!\!\!\!\!\!J_1 = \int_{-\infty}^{+\infty}\frac{\mathrm{d}\omega}{4\pi m} (\omega + \Omega){\cal J}_1(\omega + \Omega) N(\omega)\chi''_{0}(\omega),\label{eq:j1int}
\end{equation}
where we have introduced 
\begin{equation}
N(\omega) = \coth\left(\frac{\omega + \Omega}{2T_1}\right)-\coth\left(\frac{\omega}{2T_2}\right) \label{eq:ENNE}
\end{equation}
and the response function $\chi_0(t)$ (see Appendix~\ref{app:pert}) whose Fourier transform is
\begin{equation}
\chi_0(\omega)=-\frac{1}{\omega^2-\omega_0^2+i\omega\gamma_2(\omega)}\,,\label{eq:chi0}
\end{equation}
where $\gamma_2(\omega)=\int_{-\infty}^{+\infty}\mathrm{d}t e^{i\omega t}\gamma_2(t)$ is the damping kernel of the $\nu=2$ bath in Fourier space with
\begin{equation}
\gamma_{\nu}(t) \equiv \sum_{k=1}^{+\infty} \frac{c_{k,{\nu}}^2}{m_{k,{\nu}}\omega_{k,{\nu}}^2}\cos(\omega_{k,{\nu}}t)\,.
\label{eq:Gamma}
\end{equation}
Here and in the following the double prime denotes the imaginary part.
In the following we will focus only on $J_1$ and $P$, but $J_2$ can be easily obtained from the energy balance expressed by Eq.~(\ref{balance}).
The auto--correlation functions for the total power, $D_P$, and for the heat current, $D_{J_1}$, are
\begin{equation}
D_{P}=\Omega^2\int_{-\infty}^{+\infty}\frac{\mathrm{d}\omega}{4\pi m}~ {\cal J}_1(\omega+\Omega) R(\omega) \chi''_{{0}}(\omega),\label{eq:dppint}
\end{equation}
\begin{equation}
\!\!\!\!\!D_{J_1}\!\!=\!\!\int_{-\infty}^{+\infty}\!\!\frac{\mathrm{d}\omega}{4\pi m}\!(\omega +\Omega)^2 {\cal J}_1(\omega+\Omega) R(\omega)\chi''_{{0}}(\omega),\label{eq:d11int}
\end{equation}
with
\begin{equation}
\!\!R(\omega)=\coth\left(\frac{\omega+\Omega}{2T_1}\right)\coth\left(\frac{\omega}{2T_2}\right) -1\,.\label{eq:ERRE}
\end{equation}

Hereafter, we focus on the specific case of a system with a structured, Lorentzian spectral density for the dynamically coupled bath
\begin{equation}
{\mathcal J}_1(\omega)=\frac{d_1 m\gamma_1\omega}{(\omega^2-\omega_1^2)^2+\gamma_1^2\omega^2}\,,\label{eq:spden1}
\end{equation}
peaked at $\omega\approx\omega_1$, with typical broadening $\gamma_1$, and coupling strength parameterized by $d_1$. Such a spectral density offers a great versatility, i.e., exploiting different resonant conditions between external drive $\Omega$ and $\omega_1$, and is known to induce non--Markovian effects~\cite{cavaliere_prr,nm1,nm2,nm3,nm4,nm5}, a necessary condition to have a dynamical heat engine in this setup~\cite{cavaliere_prr}. In the case of a Ohmic spectral density, instead, the present quantum thermal machine would not operate as heat engine~\cite{cavaliere_prr}.
A Lorentzian spectral density can be either obtained in cavity--optomechanical systems~\cite{cavop1,cavop2} or in circuit-QED setups~\cite{cqed1,cqed2}, with the latter representing a particularly convenient solid--state platform which allows to tune dynamical couplings up to a high degree of precision~\cite{cqed3}. For the statically coupled contact we choose a Ohmic spectral density ${\mathcal J}_2(\omega)=m\gamma_2\omega$, which implies $\gamma_2(\omega)\equiv\gamma_2$ in Eq.~(\ref{eq:chi0}).

\subsection{Average thermodynamic quantities}
The quantum-thermodynamical properties of this setup have been investigated recently~\cite{cavaliere_prr}. It has been shown that it can either operate as a quantum engine or a quantum refrigerator depending on the parameters. Here we focus on the case $\gamma_{1,2}\ll\omega_0$, favourable to obtain better performances and smaller fluctuations, and on temperatures in the quantum regime $T_{\nu}<\omega_0$. In particular, we show here the case $T_2=2 T_1$, with $T_1=0.4\omega_0$.\footnote{To be consistent with the perturbative approach, $d_1$ in Eq.~(\ref{eq:spden1}) has been chosen small enough so that the peak value of $\mathcal{J}_1(\omega)$ satisfies $\mathcal{J}_1(\omega_1)\ll m\omega_0^2$.}

\begin{figure}[htbp]
    \centering
    \includegraphics[width=7.5cm,keepaspectratio]{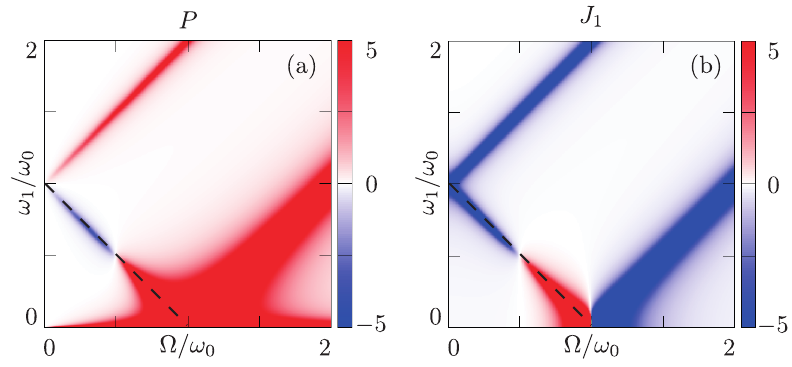}
    \caption{{\bf Energy currents and operating modes}. Panels (a) and (b) show, as a density plot, the average power $P$ and heat current $J_1$ as a function of the driving frequency $\Omega$ and the resonance frequency $\omega_1$. Both quantities are here normalized to $\gamma_2^2$. The dashed line marks the resonance condition $\omega_1=\omega_0-\Omega$. In all panels $T_1=0.4\omega_0$, $T_2=0.8\omega_0$, $\gamma_1=0.05\omega_0$, $\gamma_2=0.01\omega_0$, and $d_1=10^{-3}\omega_0^4$.}
    \label{fig:enemodes}
\end{figure}
Results for the total power $P$ and for the heat current $J_1$, obtained by numerically integrating Eqs.~(\ref{eq:pint}),(\ref{eq:j1int}) are shown as density plots in Figs.~\ref{fig:enemodes}(a,b) as a function of the driving frequency $\Omega$ and $\omega_1$. Depending on the sign of $J_{\nu}$ and $P$ different operating modes are identified (see Appendix~\ref{app:opmodes})~\cite{cavaliere_iscience}. Here we focus on the most relevant ones: Heat engine with $P<0$ and $J_1<0$ and refrigerator with $P>0$ and $J_1>0$. The best performance for these operating modes occurs, for $T_1<T_2$, when $\omega_1=\omega_0-\Omega$ is satisfied~\cite{cavaliere_prr}, see the dashed line.\footnote{The situation for $T_1>T_2$ is essentially mirrored along the axis $\omega_1=\omega_0$. In this case the best engine and refrigerator operating regimes occur along the resonance condition $\omega_1=\omega_0+\Omega$~\cite{cavaliere_prr}.}

We mention that the performance of proposed dynamical quantum heat engine has been investigated beyond the weak-coupling regime in Sec.~IV.C of Ref.~\cite{cavaliere_prr}. In that regime, we observe a broadening of the power resonance line, which appears to be detuned with respect to $\omega_1 = \omega_0 -\Omega$, and it is possible to achieve sensibly larger power outputs. However, higher efficiency is observed in the weak-coupling regime. The heat engine is lost in the limit of very strong coupling.

\subsection{Thermodynamic uncertainty relations}
It is now interesting to assess the impact of fluctuations on the performance of this setup. To this end, we use TURs~\cite{barato,cangemi, horowitz17, brandner18, agarwalla}. We are particularly interested in the engine and refrigerator regimes and thus in the TURs for $P$ and $J_1$, expressed in terms of the 
trade-off parameter $Q_P$ and $Q_{J_1}$ as~\cite{cangemi, brandner18, agarwalla}
\begin{equation}
Q_P=\dot{S}\frac{D_{P}}{P^2}\ge 2\ ;\ Q_{J_1}=\dot{S}\frac{D_{J_1}}{J_1^2}\ge 2\,.
\label{eq:QP_QJ1_def}
\end{equation}
These TURs combine energy flows, their fluctuations, and the entropy production rate in a dimensionless quantity expressing the trade-off between how the system fluctuates with respect to the quality (in terms of magnitude and degree of dissipation) of the energy flow. The lower $Q_{\mu}$ (with $\mu\in\{P,J_1\}$), the better the operation of the thermal machine~\cite{pietzonka18}. 
\begin{figure}[htbp]
    \centering
    \includegraphics[width=7.5cm,keepaspectratio]{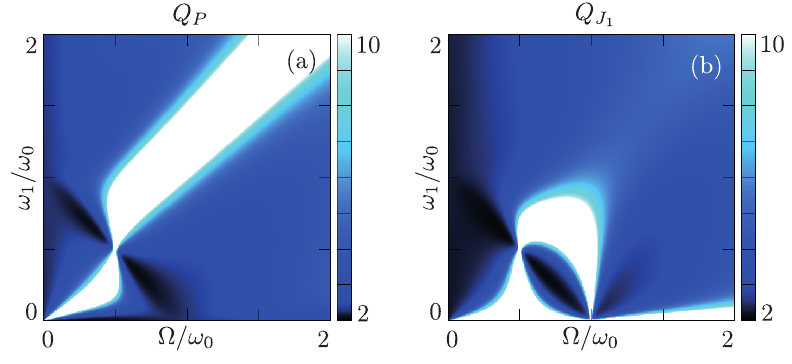}
    \caption{{\bf Quantifying fluctuations via TURs}. Density plot of $Q_P$ (panel a) and $Q_{J_1}$ (panel b) as a function of $\Omega$ and $\omega_1$. Parameters as in Fig.~\ref{fig:enemodes}.}
    \label{fig:denstur}
\end{figure}
Figure~\ref{fig:denstur} shows density plots of the numerically evaluated $Q_P$ (a) and $Q_{J_1}$ (b) as a function of $\Omega$ and $\omega_1$ and it is clear that $Q_{\mu}>2$ everywhere. Via extensive numerical investigations we can report that regardless of all parameter configurations, $Q_{\mu}<2$ is never found in this model, and thus no violation of TUR bounds~\cite{barato,cangemi,brandner18} occurs. The trade-off parameters can attain very large values, see the white regions in the plots, and even diverge -- e.g. at the crossover between the engine and the refrigerator regimes along $\omega_1=\omega_0-\Omega$ (due to $J_1$, $P$ crossing zero). Interestingly, however, they are instead particularly low (even approaching $Q_{\mu}\approx 2$) around the latter resonance line, where the best performances occur. Let us then relate $Q_{\mu}$ to the efficiency $\eta=-P/J_2$ (engine mode) or to the coefficient of performance $\mathrm{COP}=J_1/P$ (refrigerator mode), normalized to the respective Carnot limits $\eta_C=1-\frac{T_1}{T_2}$ and $\mathrm{COP}_C=\frac{T_1}{T_2-T_1}$.\footnote{For the parameters chosen here, $\eta_C=0.5$ and COP$_C$=1.}
 \begin{figure}[htbp]
    \centering
    \includegraphics[width=7.5cm,keepaspectratio]{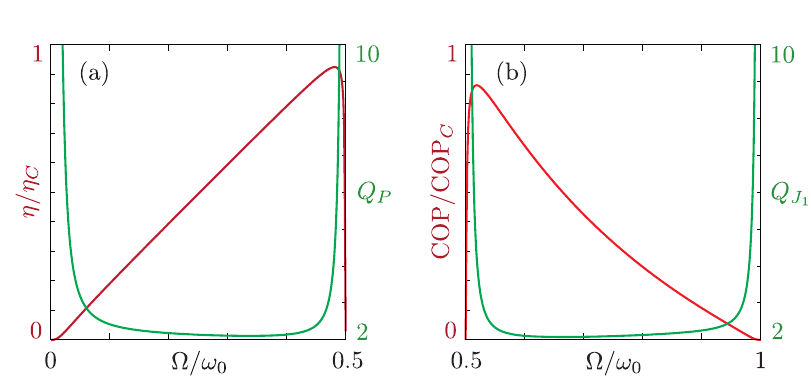}
    \caption{{\bf Efficiency vs. fluctuations}. Panel (a) shows as a red curve the efficiency, normalized to the Carnot limit, of the thermal machine operating as an engine, and $Q_P$ as a green line, as a function of $\Omega$ along the resonance line $\omega_1=\omega_0-\Omega$ -- see Fig.~\ref{fig:enemodes}. Panel (b) shows the same quantities along the same resonance, but in the regime where the machine operates as a refrigerator. Parameters as in Fig.~\ref{fig:enemodes}.}
    \label{fig:tagli}
\end{figure}
Figure~\ref{fig:tagli}(a) shows $\eta/\eta_C$ (red line) and $Q_P$ (green line) as a function of $\Omega$ within the engine region along the dashed line of Fig.~\ref{fig:enemodes}. As is clear, there is a large range of driving frequencies where $Q_P$ is very close to 2. Furthermore, when it attains its minimum value, for $\Omega\approx0.35\omega_0$, the efficiency of the engine is $\approx0.7\eta_C$. On the other hand, when the engine operates at its maximum efficiency of $\eta\approx0.95\eta_C$, $Q_P$ almost doubles. Notice also that $Q_P$ diverges at the boundaries of the engine regime and that the maximum of $\eta$ is located in close proximity of one of these edges. The situation in the refrigerator regime is qualitatively similar, and is reported in Fig.~\ref{fig:tagli}(b).

From a different perspective, the trade-off parameter $Q_P$ ($Q_{J_1}$) in Eq.~(\ref{eq:QP_QJ1_def}) expresses the performance of the heat engine (refrigerator) as a trade-off between three desiderata~\cite{pietzonka18}: (i) Finite (or even
large) average output power $-P>0$ (cooling power $J_1>0$), (ii) an efficiency $\eta$ (COP) close to the Carnot value $\eta_C$ (COP$_C$), and (iii) constancy, i.e., small fluctuations $D_{P}$ ($D_{J_1}$). Indeed, we can rewrite Eq.~(\ref{eq:QP_QJ1_def}) as
\begin{align}
& Q_{P} = - \frac{1}{T_1}\frac{D_P}{P}\left [ \frac{\eta_C}{\eta} - 1\right ]\geq 2 ,\nonumber\\
& Q_{J_1} = \frac{1}{T_2}\frac{D_{J_1}}{J_1}\left[ \frac{1}{\mathrm{COP}} - \frac{1}{\mathrm{COP}_C} \right ] \geq 2,
\end{align}
where, we recall, $T_1<T_2$. Therefore, attaining values $Q_\mu \approx 2$ means achieving a nearly optimal performance of the machine, intended as a trade-off between the above desiderata.

To gain further insight, we can exploit the fact that in the limit $\gamma_2\to 0$ one can approximate~\cite{cavaliere_prr} $\chi''_0(\omega)\approx\frac{\pi}{2\omega_0}\sum_{p=\pm 1}p\delta(\omega-p\omega_0)$, with $\delta(\omega)$ the Dirac delta. The integrals in Eqs.~(\ref{eq:pint}),(\ref{eq:j1int}),(\ref{eq:dppint}),(\ref{eq:d11int}) are then easily solved, leading to
\begin{eqnarray}
P&=&-\frac{\Omega}{8m\omega_0}\sum_{p=\pm 1}p\mathcal{J}_1(\omega_0+p\Omega)N_p ,\nonumber\\
J_1&=&\frac{1}{8m\omega_0}\sum_{p=\pm 1}(\omega_0+p\Omega)\mathcal{J}_1(\omega_0+p\Omega)N_p ,\nonumber\\
D_{P}&=&\frac{\Omega^2}{8m\omega_0}\sum_{p=\pm 1}\mathcal{J}_1(\omega_0+p\Omega)R_p ,\nonumber\\
D_{J_1}&=&\frac{1}{8m\omega_0}\sum_{p=\pm 1}(\omega_0+p\Omega)^2\mathcal{J}_1(\omega_0+p\Omega)R_p ,\nonumber
\end{eqnarray}
where $N_p=N(p\omega_0)$ and $R_p=R(p\omega_0)$ -- see Eqs.~(\ref{eq:ENNE}),(\ref{eq:ERRE}). In the regime $\gamma_1\ll\omega_0$ considered in this paper, restricting along the resonance line $\omega_1=\omega_0-\Omega$ one can safely drop all terms with $p=+1$ in the above expressions~\cite{cavaliere_prr}. Then, after simple rearrangements one can see that in this limit $Q_P\equiv Q_{J_{1}}$, with
\begin{equation}
Q_{\mu}\!=\!2\left(\frac{\omega_0}{2T_2}-\frac{\omega_0-\Omega}{2T_1}\right)\coth\left(\frac{\omega_0}{2T_2}-\frac{\omega_0-\Omega}{2T_1}\right).\nonumber
\end{equation}
This proves that, at least in the above limits, the TURs are never violated and $Q_{\mu}\geq 2$ in agreement with our numerical findings. Indeed, in this limit $Q_{\mu}=2$ precisely when $\Omega=\Omega^{*}=\omega_0\eta_C$, which is the ``turning point'' between the operation as an engine and that as a refrigerator along the resonance~\cite{cavaliere_prr}. There, at $\Omega = \Omega^{*}$, we have  $\eta\to\eta_C$ (with $P\to0$) and COP$\to$COP$_C$ (with $J_1\to0$). The net coalescence of engine and refrigerator modes is lifted at finite values of $\gamma_{\nu}$, which results in $\eta<\eta_C$ (COP$<$COP$_C$), with the maximum for $\eta$ (COP) occurring at $\Omega\lesssim\Omega^{*}$ ($\Omega\gtrsim\Omega^{*}$). Also, when $\gamma_{\nu}$ is small but non-zero the trade-off parameters are no longer identical, and develop the divergence near $\Omega^{*}$ seen in Fig.~\ref{fig:tagli}.

\section{Conclusions}\label{sec:conc}
In this paper we have analyzed the fluctuations of the heat current and power exchanged between a quantum thermal machine with a single quantum harmonic oscillator as the WM and two thermal baths at uneven temperature, in the presence of a dynamical coupling between the WM and one of the baths. Such a machine can operate either as a quantum heat engine or as refrigerator depending on the frequency of the drive and the other parameters. To understand the impact of fluctuations on such operating modes we have developed expressions for the fluctuations to leading order in the amplitude of the dynamical coupling. Such expressions allow us to numerically evaluate the trade-off parameters $Q_P$ and $Q_{J_1}$ associated to total power (in the engine case) or heat current from the cold bath (in the refrigerator case), whose magnitude is associated to the strength and impact of fluctuations. We report no violation of the lower bound proposed for these quantities in the literature. However, we have found that in typical operating regimes the trade-off parameters can reach values very close to the lower bound, implying a nearly optimal trade-off between efficiency (COP), (cooling) power output, and small fluctuations, with an efficiency of about 70\% of the Carnot limit in the case of a quantum engine. In the limit of weak damping, appropriate for the regime of parameters considered in this paper, we have also developed an analytical approximation for the trade-off parameters, which supports our findings and shows that TUR bounds cannot be violated in our model, at least for weak damping.

Despite no violation of TURs has been found, showing that the trade-off parameters can attain small values in sensible operating regimes is a first step towards the development of optimal protocols for the operation of quantum thermodynamic machines. Future developments of this work may lead to consider more complex driving protocols for the thermal machine, which could help to even increase the efficiency at the lowest values of the trade-off parameters. Also, it will be interesting to look for more complex quantum thermal machines, either in multi--terminal or multi--WM configurations. Due to their ability to perform multiple thermodynamical tasks at once~\cite{cavaliere_iscience}, assessing the impact of fluctuations can be a key issue towards their optimization. As a final outlook, further extensions of the present work may include considering squeezed baths~\cite{rossnagel_prl14,klaers_prx17,agarwalla_prb17,manzano_pre16,manzano_pre18,long_pre15,singh_pre20} to assess if and to which extent performance of the proposed quantum heat machine can be improved by non-thermal baths~\cite{Abah_2014,Alicki_2015,Niedenzu_2016,Niedenzu_2018}.

\bmhead{Acknowledgments}
L.R. and G.B. acknowledge financial support by the Julian Schwinger Foundation (Grant JSF-21-04-0001) and by INFN through the project ‘QUANTUM’. F.C. and M.S. acknowledge support by the “Dipartimento di Eccellenza MIUR 2018-2022.”

\bmhead{Data availability}
All data generated during this study have been obtained by numerically implementing the equations presented in the text using standard techniques.

\appendix

\section{Perturbative approach}\label{app:pert}
In this Appendix we derive the explicit expressions for the average thermodynamic quantities in Eqs.~(\ref{eq:pint}),(\ref{eq:j1int}), and their auto--correlation functions in Eqs.~(\ref{eq:dppint}),(\ref{eq:d11int}). We start by writing the Hamiltonian as in Eq.~(\ref{eq:totH_pert}). In the interaction picture (label $\mathrm{I}$), we freeze out the time evolution generated by $H^{(0)}$, so an observable $A$ is defined as
\begin{equation}
A_{\mathrm{I}} = e^{ i H^{(0)} (t-t_0)} A_{\mathrm{S}} e^{- i H^{(0)} (t-t_0)},
\label{eq:obs_IP}
\end{equation}
with the label $\mathrm{S}$ denoting the Schr\"{o}dinger picture. Operators in the Heisenberg (label $\mathrm{H}$) and in the interaction picture are related via
\begin{equation}
A_{\mathrm{H}}(t) = U_{\mathrm{I}}^\dagger(t,t_0) \, A_{\mathrm{I}}(t) \, U_{\mathrm{I}}(t,t_0),
\label{eq:obs_HP_IP}
\end{equation}
where
\begin{equation}
U_{\mathrm{I}}(t,t_0) = T \exp\left\{ -i \int_{t_0}^{t} {\rm d} s \Delta H_{\mathrm{I}}(s) \right\}
\label{eq:TEvol_IP_series}
\end{equation}
is the propagator in the interaction picture, with $T$ the time-ordering symbol. We are interested in the lowest-order perturbative correction with respect to the interaction strength $c_{k,1}$. Recalling that $\Delta H^{(t)} = H_{{\rm int}, 1}^{(t)}$, given in Eq.~(\ref{eq:H_int_nu}), according to Eq.~(\ref{eq:obs_IP}) we can write
\begin{align}
\Delta H_{\mathrm{I}}(t) \approx - \sum_{k=1}^{+\infty} g_1(t) c_{k,1} x_{\mathrm{I}}(t) \xi_{k,1}(t). 
\end{align}
Indeed, it is possible to show that $X_{k,1;{\mathrm{I}}} = \xi_{k,1}(t)$, with $\xi_{k,1}(t)$ defined in Eq.~(\ref{eq:xi_knu_def}).
Using Eqs.~(\ref{eq:xi_nu_def}),(\ref{eq:obs_HP_IP}) and considering $U_{\mathrm{I}}(t,t_0) \approx 1 - i \int_{t_0}^t {\rm d} s \Delta H_{\mathrm{I}}(s)$, in the Heisenberg picture the position operator of the WM can be written as
\begin{equation}
x(t)= x^{(0)}(t) + \Delta x(t),
\label{eq:x_qho_pert} 
\end{equation}
where $x^{(0)}(t) \equiv x_{\mathrm{I}}(t)$ according to Eq.~(\ref{eq:obs_IP}) and where the perturbative correction is
\begin{equation}
\!\!\!\!\!\Delta x(t) \! = \! -i \int_{t_0}^{t} {\rm d}s \,  g_1(s) \left[ x^{(0)}(s)  \xi_1(s), x^{(0)}(t)\right] .\label{eq:delta_xl_qho}
\end{equation}

\subsection{Heat current and total power}
We start by considering the average heat current $J_1$. From Eq.~(\ref{eq:J}), as detailed in Ref.~\cite{carrega_prxquantum}, we have the general, exact expression
\begin{eqnarray}
\!\!\!\!J_1&=&-\int_{\bar{t}}^{\bar{t}+{\cal T}}\frac{\mathrm{d}s}{{\cal T}}g_{1}(s)\left\langle x(s) \left[ \dot{\xi}_1(s)+\right.\right.\nonumber\\ 
&+&\left.\left.\int_{t_0}^{s} {\rm d}s'\, g_{1}(s')x(s')F_1(s-s')\right]\right\rangle\,,
\label{jj1}
\end{eqnarray}
where we have introduced
\begin{equation}
 F_1(t)=\sum_{k=1}^{+\infty}\frac{c_{k,1}^2}{m_{k,1}}\cos(\omega_{k,1} t)
\end{equation}
with Fourier transform $F_1(\omega) = 2 \omega {\cal J}_1(\omega)$. We obtain the perturbative expansion of Eq.~(\ref{jj1}),
\begin{eqnarray}
J_1\!\!&=&\!\!-\!\!\!\int_{\bar{t}}^{\bar{t}+{\cal T}}\!\!\frac{\mathrm{d}s}{{\cal T}} g_{1}(s) \bigg\{ \langle x^{(0)}(s) \dot{\xi}_1(s)\rangle + \langle\Delta x(s) \dot{\xi}_1(s)\rangle
\nonumber\\
&+&\!\!\int_{t_0}^{s} {\rm d}s'\, g_{1}(s')\langle x^{(0)}(s) x^{(0)}(s')\rangle  F_1(s-s')\bigg\},
\end{eqnarray}
by expanding $x(t)$ as in Eq.~(\ref{eq:x_qho_pert}) and retaining terms up to $O(c_{k,1}^2)$, as the latter represent the first perturbative correction to $J_1$.
We now perform the quantum average on the initial state of Eq.~(\ref{eq:initial_rho}), exploiting the fact that $\langle x^{(0)}(t) \dot{\xi}_1(t) \rangle = 0$ and that
\begin{eqnarray}
\!\!\!\!\langle \Delta x(t) \dot{\xi}_1(t)\rangle &=& -i  \int_{t_0}^{t} {\rm d}s \,g_1(s)\cdot\nonumber\\ 
\!\!\!\!&\cdot&\langle [x^{(0)}(s),x^{(0)}(t)]\rangle Z_1(s-t), 
\label{eq:q_avg_Delta_xl_xi}
\end{eqnarray}
where we have defined
\begin{equation}
Z_1(t-t')\equiv\langle\xi_1(t)\dot{\xi}_1(t')\rangle=\frac{\mathrm{d}}{\mathrm{d}t'}\mathcal{L}(t-t')\,,
\end{equation}
where the decomposition in symmetric and antisymmetric contributions
\begin{equation}
Z_1(t-t')=Z_{1,\mathrm{s}}(t-t')+Z_{1,\mathrm{a}}(t-t')\,,
\end{equation}
has been introduced, see Eq.~(\ref{corr}). Finally, we arrive at the expression for the heat current
\begin{eqnarray}
J_1&=& -\int_{\bar{t}}^{\bar{t}+{\cal T}}\frac{\mathrm{d}s}{{\cal T}}\bigg\{ \int_{t_0}^{s} {\rm d}s'g_{1}(s)g_{1}(s')F_{1}(s-s')\cdot\nonumber\\
&\cdot&\langle x^{(0)}(s) x^{(0)}(s')\rangle-i\int_{t_0}^{s} {\rm d}s'\, g_{1}(s') g_{1}(s)\cdot\nonumber\\
&\cdot&Z_1(s'-s)\langle [ x^{(0)}(s') , x^{(0)}(s) ] \rangle\bigg\}.
\end{eqnarray}

To perform the average over ${\cal T}$ we observe that the correlation function 
\begin{equation}
C(t,t')=\langle x^{(0)}(t)x^{(0)}(t')\rangle\label{ctt}
\end{equation}
only depends, in the long time limit $t\gg t_0$, on the time difference $t-t'$, i.e., in the long time limit $C(t,t')=C(t-t')$.
This allows to change variable $\tau=s-s'$, to perform the limit $t_0\to -\infty$, and thus to obtain
\begin{eqnarray}
J_1&=&-\frac{1}{2}\int_{0}^{+\infty}\!\!\!{\rm d}\tau\left\{{F_{1}(\tau)}
 \cos(\Omega \tau)C(\tau)\right.\nonumber\\ 
 \!\!\!\!&\!\!\!\!-&\!\!\!\!\left.i Z_1(-\tau)\cos(\Omega \tau)\left[C(-\tau) - C(\tau)\right]\right\}\,.~\label{eq:J1integ}
\end{eqnarray}
By direct inspection one finds that ${F_1(t)=-2i Z_{1,{\rm s}}(t)}$ and introducing the response function $\chi_0(t)\equiv i m \theta(t) \langle [ x^{(0)}(t), x^{(0)}(0)]\rangle$ (see~\cite{vignale}) with $\theta(t)$ the Heaviside step function, whose Fourier transform is given in Eq.~(\ref{eq:chi0}), it is then possible to rewrite the average heat current as in Eq.~(\ref{eq:j1int}).

Turning to the average total power of Eq.~(\ref{eq:P}), as detailed in Ref.~\cite{carrega_prxquantum}, we have the general, exact expression
\begin{eqnarray}
P&=&\int_{\bar{t}}^{\bar{t}+{\cal T}}\mathrm{d}s\bigg\{ - \dot{g}_1(s) \langle x(s) \xi_{1}(s)
\rangle+\nonumber\\
\!\!\!\!&\!\!\!\!+&\!\!\!\!\dot{g}_1(s)\left\langle x(s) \!\! \int_{t_0}^{s} \!\! {\rm d}s'\, \gamma_1(s-s') \frac{{\rm d}}{{\rm d}s'} [ g_1(s') x(s') ]\right\rangle \!\! \bigg\} .
\nonumber
\end{eqnarray}
Following the same steps of the perturbative expansion detailed above, after the quantum average we get
\begin{eqnarray}
P&=& \int_{\bar{t}}^{\bar{t}+{\cal T}}\frac{ds}{{\cal T}}\Big\{2 \int_{t_0}^{s} {\rm d}s'\,\Big[ \mathcal{L}_1(s'-s)
\dot{g}_1(s) g_1(s')\cdot\nonumber\\
&\cdot&C''(s-s')\Big]-\int_{t_0}^{s}{\rm d}s'\,\Big[  \dot{g}_1(s) g_1(s')C(s-s')\cdot\nonumber\\
&\cdot&\frac{{\rm d}}{{\rm d}s'} \gamma_1(s-s')\Big] +\gamma_1(0)\dot{g}_1(s) g_1(s) C(0)\Big\}\,.
\end{eqnarray}
Taking the average over the period of the drive, recalling Eqs.~(\ref{corr}),~(\ref{eq:ls}),~(\ref{eq:la}) and exploiting the identity $\dot{\gamma}_1(t)=-2i {\cal L}_{1,{\rm a}}(t)$ one obtains Eq.~(\ref{eq:pint}).
Notice that the above results are fully consistent with those obtained using a fully non-equilibrium Green function formalism in the perturbative regime (see Appendix E of Ref.~\cite{cavaliere_prr}).

\subsection{Correlation functions}
The auto--correlation function given in Sec.~\ref{sec:sec22} can be evaluated in the perturbative regime following steps similar to those described above. Here we briefly outline the procedure to obtain $D_{J_1}$: Starting from the definition of Eq.~(\ref{doot}), to lowest order it is sufficient to consider only the zero-th order term $x^{(0)}(t)$ of the position operator $x(t)$ in Eq.~(\ref{eq:x_qho_pert}). Then, the fluctuations of $J_1$ can be written as
\begin{eqnarray}
D_{J_1}(t)&=&\!\!\int_0^{+\infty}{\mathrm d}\tau g_{1}(t)g_1(t-\tau)\cdot\nonumber\\
\!\!&\!\!\cdot&\!\!\left\langle \{x^{(0)}(t)  \dot{\xi}_1(t),x^{(0)}(t-\tau)\dot{\xi}_1(t-\tau)\} \right\rangle\,.\nonumber
\end{eqnarray}
The quantum averages in the above expression decouple in terms of the form $\langle x^{(0)}(t)x^{(0)}(t')\rangle\equiv C(t,t')$ -- see Eq.~(\ref{ctt}) -- and in terms of the form
\begin{equation}
\langle \dot{\xi}_1(t)\dot{\xi}_1(t')\rangle=\frac{{\mathrm d}}{{\mathrm d}t}\frac{{\mathrm d}}{{\mathrm d}t'}{\cal L}_1(t-t')\,.
\end{equation}
Exploiting again the time translational invariance $C(t,t')=C(t-t')$ and following steps analogous to those discussed above, the fluctuations of $J_1$ averaged over ${\cal T}$ -- see Eq.~(\ref{corrav}) -- turn out to be Eq.~(\ref{eq:d11int}).
In complete analogy, we derive the period--averaged fluctuations of the power $P$ in Eq.~(\ref{eq:dppint}).

\section{Operating modes}
\label{app:opmodes}
\begin{figure}
    \centering
    \includegraphics[width=4.cm,keepaspectratio]{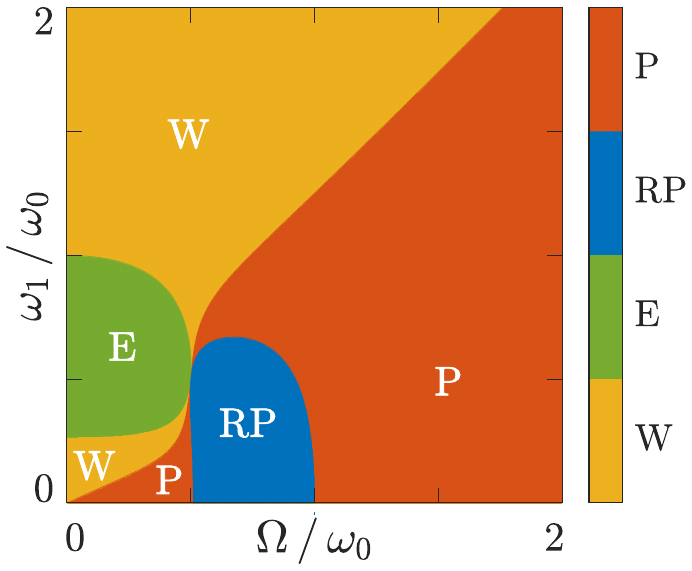}
    \caption{{\bf Operating modes}. All the possible operating modes of the quantum thermal machine discussed in main text as a function of the driving frequency $\Omega$ and the resonance frequency $\omega_1$: Heat pump (P), refrigerator--pump (RP), heat engine (E), and wasteful (W). Parameters as in Fig.~\ref{fig:enemodes}.}
    \label{fig:opmodes}
\end{figure}

The operating modes of a thermal machine where the WM is connected to two thermal baths ($\nu =1,2$) can be classified studying the sign of the heat currents $J_\nu$ and of the total power $P$. Assuming $T_1 < T_2$, the only achievable modes in a two--terminal device are the two pure ``heat engine'' ($P<0$, $J_1<0$, $J_2>0$) and ``heat pump'' modes ($P>0$, $J_1<0$, $J_2<0$), the hybrid ``refrigerator--pump'' mode ($P>0$, $J_1>0$, $J_2<0$), and the ``wasteful" mode ($P>0$, $J_1<0$, $J_2>0$)~\cite{cavaliere_iscience}. The latter is commonly labeled as ``wasteful''~\cite{hajiloo, manzano20} since in this configuration heat flows from the hot bath at $T_2$ to the cold one at $T_1$ while the WM absorbs power. We note that other works~\cite{cavaliere_prr,Buffoni19,Solfanelli20} use different names for such operating modes, e.g., ``dissipator'' for ``heat pump'' and ``accelerator'' for ``wasteful''. We also use, in the main text, ``refrigerator'' in place of ``refrigerator--pump'' for shortness, although it actually corresponds to the hybrid mode where the machine simultaneously operates as refrigerator and heat pump. Indeed, the pure ``refrigerator'' mode can not be observed in a two--terminal device~\cite{manzano20}. All the possible operating modes of the quantum thermal machine discussed in Sec.~\ref{sec:sec4} are shown in Fig.~\ref{fig:opmodes} (see also Supplemental Information in Ref.~\cite{cavaliere_iscience}). We recall that, in this setup, the bath $\nu = 1$ is the Lorentzian bath with dynamical coupling, the bath $\nu = 2$ is the Ohmic bath with static coupling, and $J_2 = -P-J_1$ from Eq.~(\ref{balance}).

\end{document}